\documentclass{ifacconf}

\usepackage{graphicx}      % include this line if your document contains figures
\usepackage{natbib}        % required for bibliography
\usepackage{amsfonts}
\usepackage{amsmath}
\usepackage{amssymb}
\usepackage{comment}
\usepackage{mathtools}
\mathtoolsset{showonlyrefs}
\usepackage[dvipsnames]{xcolor}
\newcommand{\E}{\mathbb{E}}
\newcommand{\Ghatn}{\hat{G}_N(e^{i\omega})}
\definecolor{uugreen}{HTML}{14B03D}
\definecolor{uured}{RGB}{191,45,56}
\definecolor{uublue}{RGB}{0, 106, 178}

\begin{document}
\begin{frontmatter}

\title{Privacy and Security in Network Controlled Systems via Dynamic Masking \thanksref{footnoteinfo}}

\thanks[footnoteinfo]{This work is supported by the Swedish Research Council under the grant 2018-04396 and the Swedish Foundation for Strategic Research.}

\thanks[footnoteinfo2]{These authors contributed equally.}

\author[First]{Mohamed Rasheed Abdalmoaty\thanksref{footnoteinfo2}} 
\author[Second]{Sribalaji C. Anand\thanksref{footnoteinfo2}}
\author[First]{Andr{\'e} M. H. Teixeira}

\address[First]{Department of Information Technology, Uppsala University,\\ PO Box 337, SE-75105, Uppsala, Sweden. \\
(e-mail: \{mohamed.abdalmoaty, andre.teixeira\}@it.uu.se)}
\address[Second]{Department of Electrical Engineering, Uppsala University,\\ PO Box 534, SE-75121, Uppsala, Sweden (e-mail:sribalaji.anand@angstrom.uu.se)}

\begin{abstract}                % Abstract of not more than 250 words.
% We propose some ideas of using dynamic models to ``mask" private features of the controlled systems. Suggested paper title: ``Privacy and Security in Network Controlled Systems via Dynamic Masking" 
In this paper, we propose a new architecture to enhance the privacy and security of networked control systems against malicious adversaries. We consider an adversary which first learns the system dynamics (privacy) using system identification techniques, and then performs a data injection attack (security). In particular, we consider an adversary conducting zero-dynamics attacks (ZDA) which maximizes the performance cost of the system whilst staying undetected. However, using the proposed architecture, we show that it is possible to (i) introduce significant bias in the system estimates of the adversary: thus providing privacy of the system parameters, and (ii) efficiently detect attacks when the adversary performs a ZDA using the identified system: thus providing security. Through numerical simulations, we illustrate the efficacy of the proposed architecture.
\end{abstract}

\begin{keyword}
Networked systems, Secure networked control systems, Linear systems, Privacy, Control over networks.
% {\color{red}Need to choose more keywords from \url{https://www.ifac-control.org/areas/technical-committees-and-their-scopes}}
\end{keyword}

\end{frontmatter}
%===============================================================================

\section{Introduction}
Networked control systems (NCSs) are of significant importance to society as they have become part of our daily life. Examples include the power grids \citep{singh2014stability}, and water supply networks \citep{cembrano2000optimal}. However, due to the increased use of (possibly) un-secure open communication channels in NCS, they are prone to cyber-attacks. The social and economical consequences of such cyber-attacks can be disastrous \citep{dibaji2019systems,sandberg2022secure}. Thus preventing, detecting, and mitigating such attacks is of utmost importance, and is currently an active research area with several contributions based on different approaches \citep{ferrari2021safety}. 

In the literature, there are different security concepts such as physical watermarking, moving target defense, and multiplicative watermarking \citep{chong2019tutorial}. Such security concepts focus on detecting cyber attacks. On the other hand, various privacy concepts help reduce unauthorized access to the transmitted data \citep{nekouei2019information}, thus mitigating attacks. 

However, except in a few works \citep{mukherjee2021secure}, privacy and security are considered independently. In practice, a system operator prefers privacy (of data or system properties), and in the worst case, prefers to be secure (able to detect) cyber-attacks. Thus, inspired by internal model control \citep{zhang2010advanced}, we propose an architecture for NCSs to provide a unified framework for privacy and security (which will be defined later). 

\begin{figure}
    \centering
    \includegraphics[width=8.4cm]{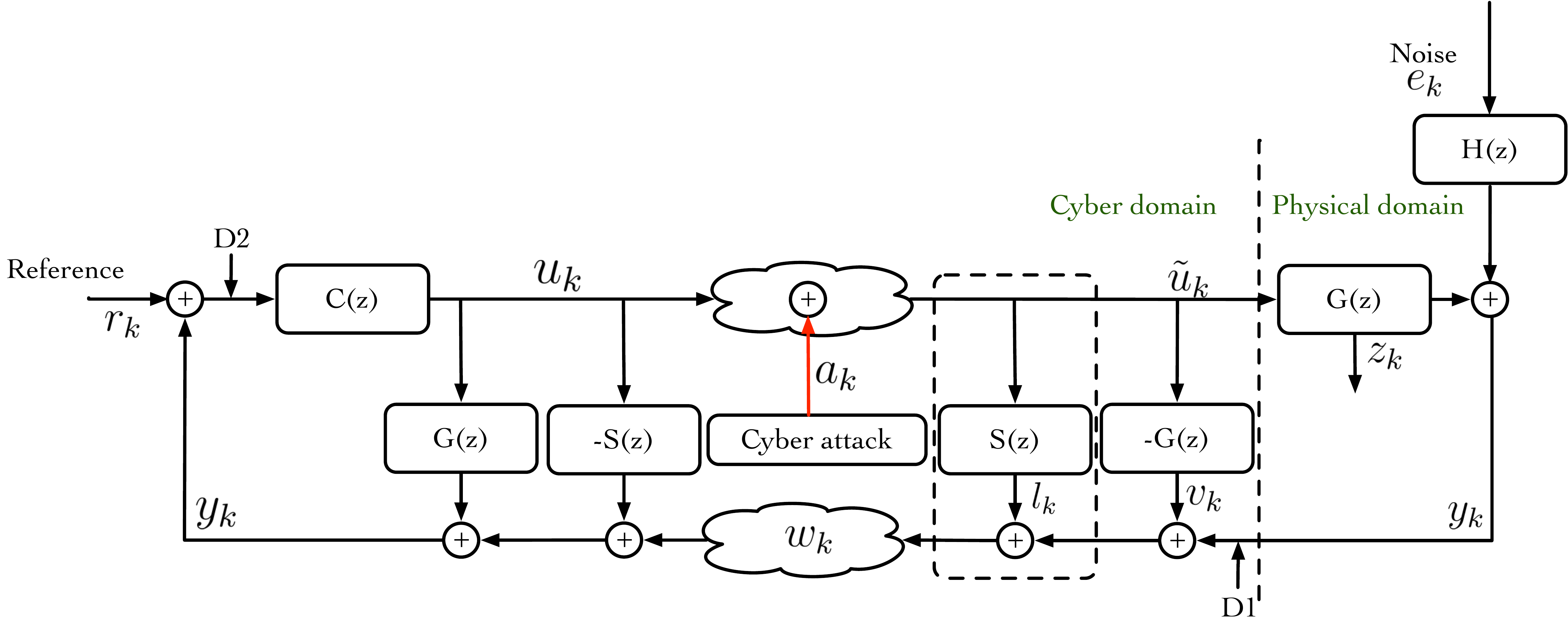}
    \caption{Proposed dynamic masking architecture. Here $z$ represents the $Z-$transform operator. $D1$ and $D2$ represent the two possible locations of the detector. The arbitrary system $S(z)$ in the dotted box is the plant seen by the adversary.}
    \label{fig1}
\end{figure}

In particular, we consider a Discrete-Time (DT) Linear Time-Invariant (LTI) description of a plant $(G)$ on one side of the network. The plant runs in parallel with simulations of $G$ and an arbitrary system ${S}$; these can be seen as being parts of a privacy filter layer or a smart sensor. On the other side of the network are the detector, the reference signal and controller, and similar simulations of $G$ and $S$ which can be thought of as being parts of a Digital Twin (DiT) \citep{barricelli2019survey}. The exact mathematical description of the system is given in the next section, while a pictorial representation is given in Fig. \ref{fig1}.  

With the architecture in Fig. \ref{fig1}, referred to as dynamic masking architecture, we consider an adversary deploying an attack in a two-step procedure (similar to \citet{mukherjee2021secure}): first learn the system dynamics, and then inject an attack which is not detected but deteriorates the system performance. The key contributions of our paper under the proposed architecture are the following:\vspace{-0.1cm}
\begin{enumerate}
    \item We propose the Mean Squared Error (MSE) between the true plant and the system learnt by the adversary as a measure of privacy. That is, we define privacy in terms of the system parameters and not in terms of the signals themselves.
    \item We show that the operator can introduce an arbitrary amount of bias into the parameters of the model identified by the adversary. In other words, the adversary will only be able to identify the arbitrary system $S$ and not the plant $G$. 
    \item We show that the attack performance deteriorates: the attack is effectively detected under some conditions on $S$. 
\end{enumerate}

Our approach is related to other works in the literature such as watermarking in that we require time-synchronization between the plant side and the controller side, and use dynamical filters in the cyber domain. However, to our knowledge, the use of dynamic watermarking for privacy was not considered before. Additionally, we do not require the invertibility of the filters used. Our work is also related to the 2-way coding \citep{fang2019two}. However, they do not provide a measure of privacy.
% use of two-way coding for privacy was not studied.

The remainder of this paper is organized as follows: we formulate our problem in Section \ref{sec:problem_formulation}. Using the proposed system architecture, we discuss the privacy aspect in Section \ref{sec:Privacy} and the security aspect in Section \ref{sec:Security}. We depict the efficacy of the proposed architecture in Section \ref{sec:NE}. Section \ref{sec:Conclusion} concludes the paper and provides avenues for future research. 

\textit{Notation:} Throughout this article, $\mathbb{R}, \mathbb{R}^{+}, \mathbb{Z}$ and $\mathbb{Z}^{+}$ represent the set of real numbers, positive real numbers, integers and non-negative integers respectively. Let $x: \mathbb{Z} \to \mathbb{R}^n$ be a discrete-time signal with $x_k$ as the value of the signal $x$ at the time step $k \in \mathbb{Z}$. Let the time horizon be $[0,N]=\{ k \in \mathbb{Z}^+|\; 0 \leq k\leq N \}$. The $\ell_2$-norm of $x$ over the horizon $[0,N]$ is represented as $|| x ||_{\ell_2, [0,N]}^2 \triangleq \sum_{k=0}^{N} x[k]^Tx[k]$. Let the space of square-summable signals be defined as $\ell_2 \triangleq \{ x: \mathbb{Z}^+ \to \mathbb{R}^n |\; ||x||^2_{\ell_2, [0,\infty]} < \infty\}$ and the extended signal space be defined as $\ell_{2e} \triangleq \{ x: \mathbb{Z}^+ \to \mathbb{R}^n | \;||x||^2_{[0,N]} < \infty, \forall N \in \mathbb{Z}^+ \}$. 

\section{Problem Background}\label{sec:problem_formulation}
We first describe the plant, the detector, the controller, and the system $S$. Let $q$ be the shift operator. Then, we describe the \textit{true} plant as
\begin{align}
y_k &= G_\circ(q) \tilde{u}_k + H_0(q) e_k\label{eq:plant}\\
z_k &= M_\circ(q) \tilde{u}_k 
\end{align}
where $G_\circ$, and $H_\circ$ denote the transfer functions, in the shift operator form, of the plant and the noise, respectively. Here $y \in \mathbb{R}^{n_p}$ is the measurement output of the plant, $u \in \mathbb{R}^{n_m}$ is the control input applied to the plant, $e \in \mathbb{R}^{n_p} \sim (0,Q)$ is the zero-mean process noise, and $z$ is the virtual performance output of the plant. The performance of the plant is said to be good only if the energy of the performance output $\Vert z_k \Vert_{\ell_2,[0,T]}^2$ for any time horizon $[0,\;T],T \in \mathbb{Z}^+$ is low. Such performance criterion is similar to that used in linear quadratic control. Next, we describe the DiT,
\begin{equation}\label{eq:DT}
v_k = -G_\circ(q) \tilde{u}_k.
\end{equation}
We then make the following assumption
\begin{assum}
We assume that the plant $G_\circ$ is controllable and observable. $\hfill \triangleleft$
\end{assum}

We next describe the system $S$, which will henceforth be called as \textit{cipher plant}\footnote{We adopt the nomenclature from cryptography where the \textit{plain} text ($y_k$) is encrypted as a \textit{cipher} text},
\begin{equation}\label{eq:S}
l_k = S(q) \tilde{u}_k
\end{equation}
where $l \in \mathbb{R}^{n_p}$ is the output of $S$. The signal transmitted over the communication channel then becomes
\begin{equation}\label{eq:commun}
    w_k:= y_k - v_k + l_k = S(q) \tilde{u}_k + H_0(q) e_k.
\end{equation}
Since the digital twin and the model $S$ are in the cyber domain, they do not have any measurement noise.

The controller and the detector are described as
\begin{align}
    u_k&=C_\circ(q)\begin{bmatrix} {y}_k^T & r_k^T\end{bmatrix}^T\label{eq:C}\\
    d_k&=D_\circ(q){y}_k\label{eq:D}
    % d_k&=D_\circ(q)\begin{bmatrix} {y}_k^T & r_k^T & u_k^T\end{bmatrix}^T\label{eq:D}
\end{align}
where $u$ is the control signal generated by the controller, $r \in \mathbb{R}^{n_p}$ is the reference signal, and $d_k \in \mathbb{R}^{n_p}$ is the output produced by the detector. The controller is designed such that the output $y$ follows the reference signal $r$ and the closed-loop system is asymptotically stable.

In the sequel, a detector is located at the plant output ($D1$). We justify the detector placement at the plant output due to the increased use of smart sensors in control \citep{walsh2001scheduling}. In the general NCS literature, the detector is located at $D2$. We later provide results on how the placements affect the results on attack detection. The detector raises an alarm if the energy of the detection output $d_k$ exceeds a predefined threshold: $\epsilon_r$. Such approaches are common in fault detection literature \citep{isermann2006fault}. The difference between $u$ and $\tilde{u}$ will be explained later. For now, they are assumed to be equal.

Observe that the signal sent by the plant is \eqref{eq:commun}. This is simply the output of the \textit{cipher} plant with some noise. However, the controller uses the signal ${y}$. This is valid because, in the absence of any cyber-attacks, due to the architecture, we can compensate for the \textit{cipher} plant $S$ and the digital twin and reconstruct the output signal $y$. 

Note that the control signal generated by the controller $u$ can be different from the control signal received by the plant $\tilde{u}$. This is because we consider an adversary capable of inserting malicious data packets. We next describe the resources and objectives of the adversary. 

\begin{rem}
Given a shift operator model ($G_\circ(q)$ for example), there always exists an equivalent state-space form whose matrices are given by $(A_{G}, B_{G}, C_{G}, D_{G})$. Thus, we use the operator form and the state-space form interchangeably throughout this paper. However, we shall indicate whenever the initial conditions are important for the analysis of security and privacy. $\hfill \triangleleft$
\end{rem}
\begin{rem}
Without loss of generality, in the rest of the paper, the performance output $z_k$ is considered to be the full state vector of the plant. $\hfill \triangleleft$
\end{rem}

\subsection{Adversary description}
In the communication channel between the controller and the plant, we consider an adversary injecting false data into the actuator of the plant. Next, we discuss the resources the adversary has \citep{teixeira2015secure}.

\subsubsection{Disclosure and disruption resources:} The adversary can eavesdrop on the communication channels and collect data. We represent the data available to the adversary at any time instant $i \in \mathbb{Z}^{+}$ as 
\begin{equation}
    \mathcal{I}_i := \cup_{t=0}^{i}\{w_k,u_k\}
\end{equation}
The adversary can also inject data into the control communication channels. This is represented by: 
\begin{equation}
    \tilde{u}_k = u_k + a_{k}
\end{equation}
where $a_{k}$ is the data injected by the adversary. Here $a \in \mathcal{L}_{2e}$ since the $\mathcal{L}_{2e}$ space allows us to study a wider class of attack signals than  $\mathcal{L}_{2}$.

\begin{figure}
    \centering
    \includegraphics[width=8.4cm]{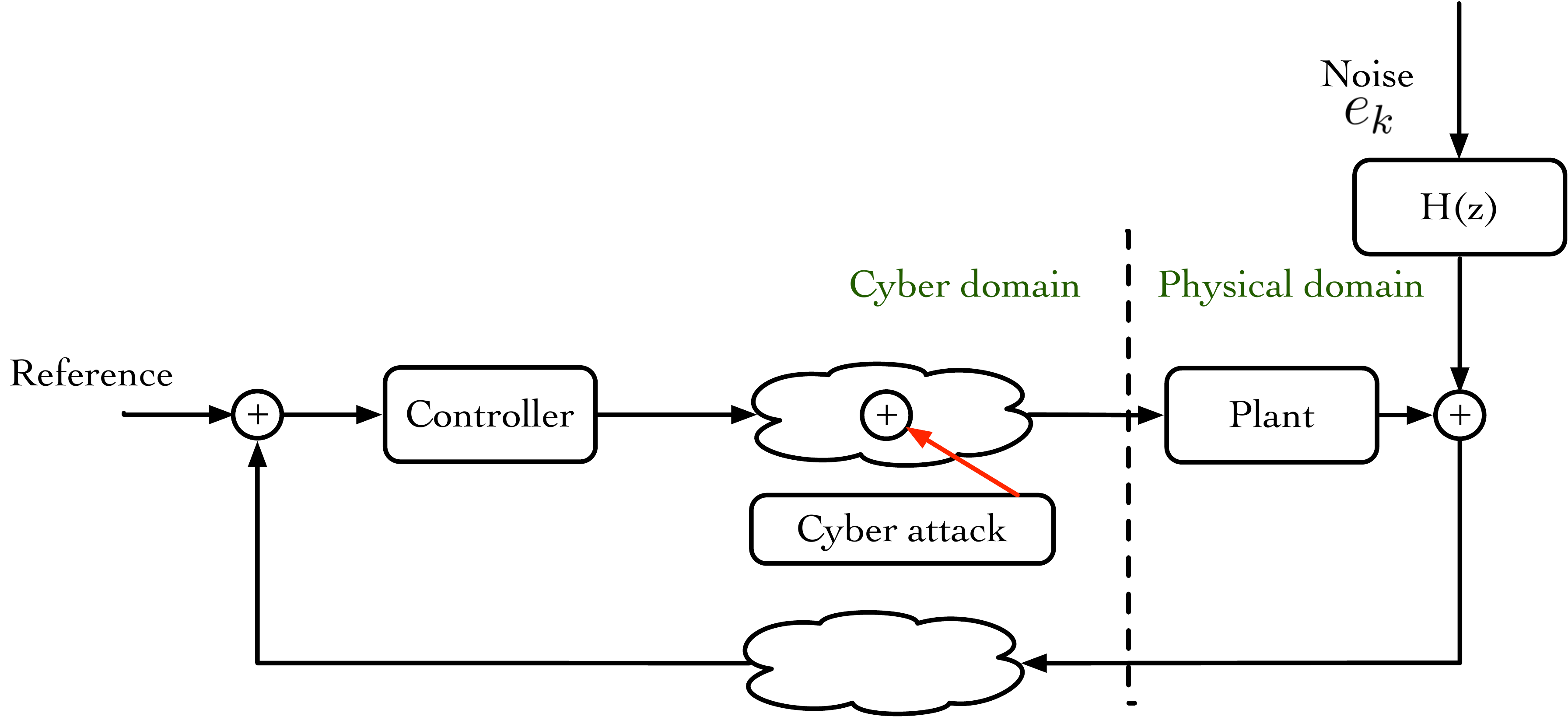}
    \caption{Architecture believed by the adversary.}
    \label{fig2}
\end{figure}

\subsubsection{Plant knowledge:} The adversary knows the order of the plant and the presence of at least one zero in the plant. Except for this, the adversary does not know anything about the NCS. They however devise an estimator of the plant $\mathcal{I}_i \mapsto \hat{G}$, 
 from the disclosed data, according to the believed model specifications in  Fig.~\ref{fig2}. We explain in the next section, how the dynamic masking architecture distorts $\hat{G}$. 

\subsubsection{Attack goals and constraints:}
The adversary aims at deteriorating the system's performance while remaining undetected. Hence, the adversary injects attack signals to maximize the energy of the states of the system $\hat{G}$ output while keeping the energy of the detection output lower than $\epsilon_r$. The aim of maximizing the energy of the states is to consequently maximize the energy of the output $z$ (this objective is contradictory to the plant).

\subsection{Problem Formulation}
Many methods can be used to construct stealthy data injection attacks \citep{fotiadis2020constrained,anand2021stealthy}, and in this paper, we consider zero-dynamics attacks (ZDA). Thus, to help us define security and privacy for the operator, we make the following assumptions.
\begin{assum}\label{ass_un_zero}
$G$ has at least one zero. $\hfill \triangleleft$. 
\end{assum}

As depicted in \citep{teixeira2015secure}, a ZDA is constructed based on the location of the zero. Thus, we adopt the following definition of privacy in our paper.
\smallskip

\begin{defn}\label{defn:privacy}(Privacy of a property of $G$ with respect to an adversary)
Let $\psi_G$ be any property of $G$ (e.g., a zero of $G$). Then $\psi_G$ is said to be $\delta$-private with some $\delta>0$, if the estimator $\widehat{\psi_G}= \psi_{\hat{G}}$ of $\psi_G$ used by the adversary, based on disclosed data $\mathcal{I}_i$, is such that $\mathbb{E}\|{\psi}_{\hat{G}} - \psi_G\|^2\geq \delta$.  $\hfill \triangleleft$
\end{defn}
%\footnote{i.e., excluding constant estimators}

According to the above definition, if $\psi_G$ is $\delta$-private, there is no way that the adversary can recover its exact true value based on  $\hat{G}(\mathcal{I}_i)$, even when $i\to \infty$. Note that $\delta$-privacy is to be established for the particular estimator (or class of possible estimators) used by the adversary. It is implicitly assumed that any used estimator possesses well-defined first- and second-order moments with respect to the underlying probability measure of the disturbances/noises. The key to establishing privacy is then to introduce bias into the adversary's inference procedure, via the dynamic masking architecture, as explained in Section \ref{sec:Privacy}.

As previously described, we consider a detector that raises an alarm when $\Vert d_k \Vert_{\ell_2}^2 > \epsilon_r$. Then, we adopt the following definition of security:
\begin{defn}[Security of the NCS]\label{defn:security}
The closed-loop NCS is said to be secure if one of the following holds in the presence of a ZDA:
\begin{enumerate}
    \item $\Vert d_k \Vert_{\ell_2}^2 > \epsilon_r$ if the performance deterioration is unbounded. 
    \item The performance deterioration is bounded. $\hfill \triangleleft$
\end{enumerate}
\end{defn}
Then we consider the following problem in this paper: 
\begin{prob}
Show that the dynamic masking NCS architecture proposed in Fig. \ref{fig1} provides privacy and security with respect to Definition \ref{defn:privacy} and Definition \ref{defn:security}. $\hfill \triangleleft$ 
\end{prob}
In the next section, we first show how the system architecture proposed in Fig. \ref{fig1} provides privacy.

\section{Privacy under the proposed architecture}\label{sec:Privacy}
\subsection{Privacy concept}
We define privacy using Definition \ref{defn:privacy}, so that a privacy leakage is understood as the ability of the attacker to infer a property $\psi_G$ of the dynamics of the controlled system from a set of measurements of the transmitted signals over the network. This is similar to the work in \cite{alisic2020ensuring}, where privacy was defined as the minimum variance
of the estimator. Instead, here, we propose the use of the MSE as a measure of privacy. Recall that the MSE can be decomposed into two terms: the squared bias and the variance. As a privacy measure, it is thus a generalization of the variance/Fisher information measure. We will be manipulating the bias instead of the variance. For the setup of NCSs, this is possible via
 the particular dynamic masking architecture in Figure \ref{fig1}.

We consider a case where  1) the adversary has access to a  set of measurements collected via eavesdropping, 2) the adversary knows the correct model structure (number of poles and zeros) and is using a consistent estimation method (with respect to the data generating mechanism) to learn the dynamics of the model. Notice that consistency is a very weak asymptotic property that is required from any sensible estimator.

\subsection{Data generating mechanisms}
As explained in the previous section, the dynamic masking architecture changes the signals transmitted over the communication channel. While the input signal is transmitted without a change, a new signal $w$ is transmitted in lieu of the plant output. This distorts the adversary's perspective of the system according to the following proposition. 

\begin{prop} Consider the dynamic masking architecture in Figure \ref{fig1}. The data generating mechanisms from the point view of the attacker with data $\mathcal{I}_i$ are as follows
\begin{itemize}
\item plant side: transfer function from $u$ to $w$
\[
\bar{G}(z) =  S(z)
\]
\item controller side: transfer function from $w$ to $u$
\[
\bar{C}(z) =  (I- C_\circ(z)(G_\circ(z) - S(z))^{-1}C_\circ(z)
\]
with the same reference signal $r$.
\end{itemize}
\end{prop}
\begin{pf}
The result is established by straightforward manipulations which are omitted here.
\end{pf}

\subsection{Bias analysis}

The goal of this part is to characterize the bias in the estimated model when a consistent estimator such as the maximum-likelihood estimator or a prediction error method estimator is used.

Associate the complex frequency function
\begin{equation}
S(e^{i\omega})  = \sum_{k=1}^\infty s_k e^{-ki\omega}, \qquad -\pi \leq \omega \leq \pi,
\end{equation}
with $i = \sqrt{-1}$, to the data generating mechanism on the plant side. Let
\[
\Ghatn = G(e^{iw}; \hat{\theta}_N)
\]
be a model of the system estimated by the adversary based on data set $\mathcal{I}_N$ of size $N$, by estimating a parameter vector $\theta$. The bias of the estimated model with respect to the data-generating mechanism is defined as
\begin{equation}
B_N(\omega) \triangleq  \E[\Ghatn] - S(e^{i\omega})
\end{equation}
and the variance is defined as
\begin{equation}
 P_N(\omega) \triangleq    \E\left[|\Ghatn -  \E[\Ghatn]  |^2\right]
\end{equation}

If the estimator used by the attacker is unbiased for the data generating mechanism, i.e., $B_N(\omega)=0 \;\forall \omega$, it holds that the MSE of the estimated model with respect to the true system is
\begin{equation}
\mathbb{E}\left[|\Ghatn -   G_\circ(e^{i\omega}) |^2\right] =  |S(e^{i\omega}) - G_\circ(e^{i\omega})| ^2 +  P_N(\omega)
\end{equation}
Otherwise, it holds that
\begin{equation}
\label{eq:finite_MSE}
\begin{aligned}
\mathbb{E}\left[|\Ghatn\right.& -    \left.G_\circ(e^{i\omega}) |^2\right] \geq \\
&\left| B_N^2(\omega) - |S(e^{i\omega}) - G_\circ(e^{i\omega})| ^2\right| + P_N(\omega)
\end{aligned}
\end{equation}
In either case, the bias is directly controlled by $S$. 
Now, suppose the attacker is using an optimal prediction error framework with a quadratic cost function\footnote{this is chosen to simplify the exposition. Notice that in that case, the optimal prediction error method coincides with the maximum likelihood when all disturbances/noise follow Gaussian distributions}  \citep{Ljung1999}, to construct an estimator of $\theta$. Then, it is well-known that, under certain mild conditions on the data and the model parameterization, the  criterion function converges to the asymptotic criterion function
\begin{equation}\label{eq:asymptotic_cost}
\bar{V}(\theta) = \frac{1}{2\pi} \int_{-\pi}^\pi \Phi_\varepsilon(\omega, \theta) d\omega
\end{equation}
where $\Phi_\varepsilon(\omega, \theta)$ is the spectrum of the parameterized prediction error, and the estimator $\hat{\theta}_N = \hat{\theta}(\mathcal{I}_N)$ converges almost surely to $\theta^\ast$, the minimizer of \eqref{eq:asymptotic_cost} over a compact set $\Theta$.

Due to linearity of the controller, the spectrum of the input signal $u$ can be decomposed as
\[
\Phi(\omega) = \Phi_u^r(\omega) + \Phi_u^e(\omega)
\]
where $\Phi_u^r(\omega) $ is the part originating from the reference signal, and $\Phi_u^e(\omega)$ is the part originating from the noise.\smallskip

\begin{thm}[\cite{Ljung1999}]\label{thm:convergence}
\begin{equation}
\hat{\theta}_N \to \arg\min_\theta \bar{V}_1(\theta) + \bar{V}_2(\theta) \quad \text{almost surely}
\end{equation}
\[
\bar{V}_1(\theta) = \int_{-\pi}^\pi |S(e^{i\omega)}) - \hat{G}(e^{i\omega}; \theta) + \Pi(e^{i\omega},\theta)|^2 \frac{\Phi_u(\omega)}{|H(e^{i\omega}, \theta)|^2} d\omega
\]
\[
\bar{V}_2(\theta) = \lambda_\circ \int_{-\pi}^\pi \frac{|H_\circ(e^{i\omega}) -H(e^{i\omega}, \theta)|^2}{|H(e^{i\omega}, \theta)|^2} \frac{\Phi_u^r(\omega)}{\Phi_u(\omega)} d\omega
\]
\[
\Pi(e^{i\omega},\theta) = \frac{\lambda_\circ}{\Phi_u(e^{i\omega})} \frac{\Phi_u^e(e^{i\omega})}{\Phi_u(e^{i\omega})} |H_\circ(e^{i\omega})-H(e^{i\omega}, \theta)|^2
\]
\end{thm}

It is easy to see that $\Pi(e^{i\omega},\theta)$ in Theorem \ref{thm:convergence} will be identically zero if the noise model coincides with the true one. This could be achieved with flexible noise models and then $\bar{V}_2 =0$. If $G$ and $H$ are independently parameterized, the asymptotic estimate becomes the minimizer of
\[
\bar{V}_1(\theta) = \int_{-\pi}^\pi |S(e^{i\omega}) - \hat{G}(e^{i\omega}; \theta)|^2 \frac{\Phi_u(\omega)}{|H_\circ(e^{i\omega})|^2} d\omega
\]
Let us assume that the adversary either has a correct/flexible noise structure, or parameterizes $G$ and $H$ independently. 
Then, we get that asymptotically in the data size
\[
\hat{G}(e^{i\omega}; \theta^\ast) = S(e^{i\omega}) \quad \forall \omega \in [-\pi, \pi]
\]
Namely, the identified model is consistent for the data-generating model $S$.

The main idea of this contribution is to make $S$ different from $G_\circ$, and thus get that
\[
\hat{G}(e^{i\omega}; \theta^\ast) \neq G_\circ(e^{i\omega}).
\]

Notice that the distribution of the bias over the frequency $\omega$ is controlled by the system operator's choice of the filter $S$. And thus, by tuning $S$ one can achieve a desired lower bound for the achievable MSE (also for the finite data case; see \eqref{eq:finite_MSE}). 
Privacy is then achieved according to Definition \ref{defn:privacy}. No matter how long data sequences the attacker uses to estimate the dynamics, the obtained estimates will be biased and the error will be lower bounded by an MSE which is a function of $S$.

\begin{rem}
    The bias analysis provided above is nonparametric in the sense that it is defined for the transfer functions. Its translation into a parametric one is straightforward under the assumption that $S$ and $G_\circ$ have the same number of zeros and poles.
\end{rem}

\begin{exmp}
Suppose that $G(z) = \frac{z-1.1}{(z-0.2)(z-0.5)}$, and let $S(z) = \frac{z-(1.1+\delta)}{(z-0.2)(z-0.5)}$ with $\delta> 0$. Then, the non-minimum phase zero of $G$ is $\delta$-private for any unbiased estimator of the zero.
\end{exmp}

\begin{rem}
Observe that even in a case where the adversary knows the true architecture in use (Fig.~\ref{fig1}), and successfully identified the cipher plant, it remains impossible to recover $G$ solely based on disclosed data: the plant is dynamically masked, and is not identifiable via $\mathcal{I}_i$ regardless of its parameterization.   
\end{rem}

\section{Security under the proposed architecture}\label{sec:Security}
In this section, we describe how the proposed architecture provides security in terms of Definition \ref{defn:security}. In the previous section, we showed that the adversary will obtain a biased estimate of $G$, which is $S$. Thus in this section, we make the following assumption
\begin{assum}\label{ass:full_know}
The adversary knows $S$ and thinks that $S$ is the plant model. $\hfill \triangleleft$
\end{assum}

We also ignore the noise $e_k$ in this section. We assume that the threshold is designed such that the noise can never trigger an alarm. There are works in the literature that design attacks that cannot raise an alarm since they are \textit{hidden} in the noise \citep{li2019design}. Such as approach is outside the scope of this paper. With these assumptions, we derive the conditions under which the ZDA is detected.

\subsection{Zero-dynamics attack}
Before describing the construction of a ZDA, we define zero-dynamics of a system $\Sigma$ next.
\begin{defn}[ZDA \citep{teixeira2015secure}]
Given a system $\Sigma$ with the state-space matrices $(A_{\Sigma}, B_{\Sigma}, C_{\Sigma}, D_{\Sigma})$, the ZDA are a class of data injection attacks, which yield the output of $\Sigma$ identically zero. The attack is of the form 
\begin{equation}\label{ZDA:attack}
    a_{k}=g\beta^k,
\end{equation}
with $g$ and $\beta$ satisfying the following equation
\begin{equation}\label{ZDA}
\begin{bmatrix}
\beta I-A_{\Sigma} & -B_{\Sigma}\\
C_{\Sigma} & D_{\Sigma}
\end{bmatrix}    
\begin{bmatrix}
x_0\\g
\end{bmatrix}=
\begin{bmatrix}
0\\0
\end{bmatrix},
\end{equation}
in which $x_0$ is the initial condition of the system $\Sigma$. $\hfill \triangleleft$
\end{defn}

We also say that the attack $a$ lies in the output nulling space of $\Sigma$. Now, given that the adversary has perfect knowledge about $S$ and the architecture in Fig. \ref{fig2}, the adversary injects an attack which is the zero-dynamics of $S$. 
This is a strategic attack since it does not raise an alarm since $d_k =0,\;\forall k\in \mathbb{R}^{+}$, and the states of $S$ will diverge: making the performance deterioration unbounded. For clarity, the attack vector is of the form $a_{k}=g_s\beta_s^k$, where $\beta_s$ and $g_s$ are the zero and the input directions of $S$ respectively, and they satisfy 
\begin{equation}
\begin{bmatrix}
\beta_sI-A_{S} & -B_{S}\\
C_{S} & D_{S}
\end{bmatrix}    
\begin{bmatrix}
x_{0s}\\g_s
\end{bmatrix}=
\begin{bmatrix}
0\\0
\end{bmatrix},
\end{equation}

And, as we can see, the ZDA is dependent on the initial conditions of the plant. However, the adversary can drive the plant to the necessary initial conditions $x_0$ and then initiate a ZDA. Some works focus on a more detailed analysis on the effects of non-zero initial conditions on the stealthiness of the ZDA \citep{teixeira2012revealing}.

During the deployment of the attack, the attack might be easily detected if the reference changes. This is because the reference signal might interfere with the initial conditions necessary for the ZDA. To avoid this triviality, we make the following assumption
\begin{assum}
During the attack, $r_k \triangleq 0,\forall k\in \mathbb{R}^{+}$. Also, it is known to the adversary that $r_k \triangleq 0$ $\hfill \triangleleft$
\end{assum}
The above argument also highlights the necessity of ignoring the noise. Next, we describe how the ZDA is detected with the help of architecture in Fig. \ref{fig1}. 
\subsection{Detectability conditions}
We first present the detectability results of ZDA corresponding to an unstable zero.
\begin{thm}[Sufficient detectability conditions]\label{thm:detect}
Let $|\beta_s|\\>1$. Then it holds that $||d_k||_{\ell_2}^2 >\epsilon_r$ for the architecture in Fig. \ref{fig1} if, the unstable zeros of $S$ are not the unstable zeros of $G$.  
\end{thm}
\begin{pf}
The ZDA generated by the adversary lies in the output nulling space of $S$. The states of $S$ grow exponentially since the attack is exponential. However, in architecture Fig. \ref{fig1}, the attack passes through the plant $G$ whose output is fed to the detector. Then the attack makes the output of the plant $G$ grow exponentially: this is true since the attack does not correspond to a ZDA of $G$. This concludes the proof. $\hfill \blacksquare$
\end{pf}

The increase in the performance energy is unbounded because the ZDA corresponds to an unstable zero \citep{teixeira2015secure}. However, we showed in Theorem \ref{thm:detect} that the architecture in Fig. \ref{fig1} can provide security (with respect to (i) in Definition \ref{defn:security}), under some conditions. 

Theorem \ref{thm:detect} does not generalize to stable ZDA because the attack corresponding to a stable ZDA decays exponentially (since $|\beta_s|\leq 1$), and so does the output of $G$. Thus, if the threshold is sufficiently large, the attack is undetected. However, the increase in performance energy is bounded \citep{teixeira2015secure}. Thus, security is guaranteed with respect to (ii) in Definition \ref{defn:security}).

Until now, we considered the detector at the plant output ($D1$). This was justified by the increased use of smart sensors. However, one could also opt for the traditional architecture where the detector is present at the controller $(D2)$ with access to $\tilde{y}$ and ${u}$. Then the detector can be assumed to be a Kalman filter-based detector (see (3) in \citep{teixeira2015secure}). Now we show how the dynamic masking architecture proposed in Fig. \ref{fig1} can be used to provide security in terms of Definition \ref{defn:security}. Then, we state the following result which immediately follows from the fact that the adversary will only estimate $S$.

\begin{lem}
Let $S$ only have stable zeros. Then the increase in performance energy under a ZDA is bounded.$\square$
\end{lem}

Although the results in this section provide a way to enhance attack detection, they do not provide a general design guideline for $S$. This is left for future work but we briefly comment on the design. We want to \textit{trick} the adversary into believing that the data correspond to the plant. This can be partly achieved by setting the poles to be equal for $S$ and $G$. An additional condition is for $S$ to be internally stable. 

Thus in this section, we showed that the dynamic masking architecture proposed in Fig \ref{fig1} provides security in terms of Definition \ref{defn:security} against ZDA by either making the attack detectable or by making the performance deterioration bounded. We next depict the results of this paper through numerical examples.

\section{Numerical Example}\label{sec:NE}
\begin{exmp}
Consider the system described in Section \ref{sec:problem_formulation} with the following DT transfer functions
\begin{align}
    G(z)&=\frac{z-1.1}{(z-0.2)(z-0.5)}\\
\end{align}
\begin{align}
    C(z)&=-0.1\big(1+\frac{0.7}{z-1}\big)\\
    S(z)&=\frac{z-1.3}{(z-0.2)(z-0.5)}
\end{align}
where $z$ is the $Z$ transform operator as mentioned earlier. The detector is at $D1$. Here we chose $S(z)$ such that they have the same poles as $G(z)$ whereas the unstable zeros are different. All systems are assumed to be at zero initial condition. The sampling time of the system is $1 \mbox{sec}$. The controller is designed such that the output tracks the input for the reference signals of interest. Now we describe the aspects of privacy and security. 

\textit{Privacy:} The adversary first learns the plant by collecting data over the network $(\mathcal{I}_i)$. We constructed the system architecture in \textit{Simulink}. The data is collected for $7000 \mbox{sec}$. Then we performed system identification using the standard identification techniques \cite{Ljung1999} In particular, we used the \textit{ssest} command in \textit{MATLAB}. The collected data is compared with the response of the identified plant. This comparison is shown in Fig. \ref{fig:compare_ID}. The transfer function of $\hat{S}$ is given below
\begin{equation}
    \hat{S} = \frac{z-1.29}{(z-0.21)(z-0.49)}
\end{equation}
The Bode plot of $G, S$ and $\hat{S}$ is shown in Figure \ref{fig:bode}. Now three observations are in order. Firstly, the identified system is very close to $S$ which can be seen from the bode plot, the transfer function, and Fig. \ref{fig:compare_ID}. Secondly, the adversary has a bias in its plant knowledge as described in Section \ref{sec:Privacy} which can also be seen as the difference in Bode between $\hat{S}$ and $G$. And as mentioned before, this bias can be increased by altering $S(z)$. Thirdly, the zero of the identified plant is very different from that of the plant. Thus, according to Definition \ref{defn:privacy}, privacy is achieved: in this case, $\psi_G \triangleq \text{zero}(G(z))$ is asymptotically $0.2$-private.

\textit{Security:} Now we follow Assumption \ref{ass:full_know} and construct an attack vector for the system $S$ according to Definition \ref{ZDA}. The attack is of the form $a_k = -2.7 \times 1.3^k$ with $x_{0s}^T = - \begin{bmatrix} 0.94, & 0.19\end{bmatrix}$. The attack vector $a_k$, the output of $S, l_k$, and the states of $S$ are shown in Fig. \ref{fig:attack}. As we can see, the attack is successful in the sense that the output is always zero but the states diverge. 

According to Theorem \ref{thm:detect}, the attack should be detected with the detector at $D1$ since the unstable zeros of $S$ and $G$ are different. We can see from Fig. \ref{fig:out_g} that the output energy of $G$ increases implying detection. 
\begin{figure}
    \centering
    \includegraphics[width=8.4cm]{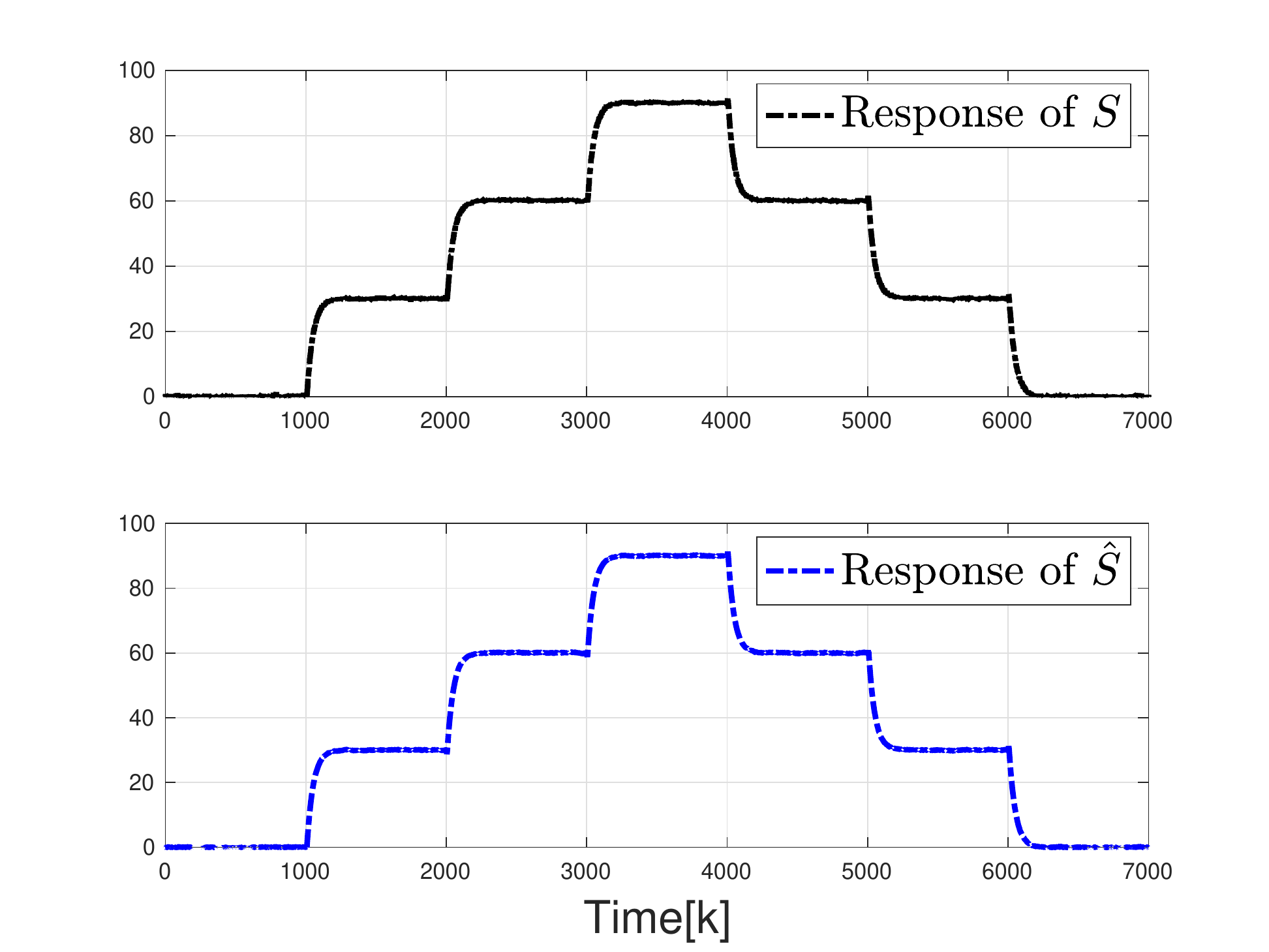}
    \caption{Comparison of the collected data $w_k$ with the response of $\hat{S}$ to $u_k$.}
    \label{fig:compare_ID}
\end{figure}
\begin{figure}
    \centering
    \includegraphics[width=8.4cm]{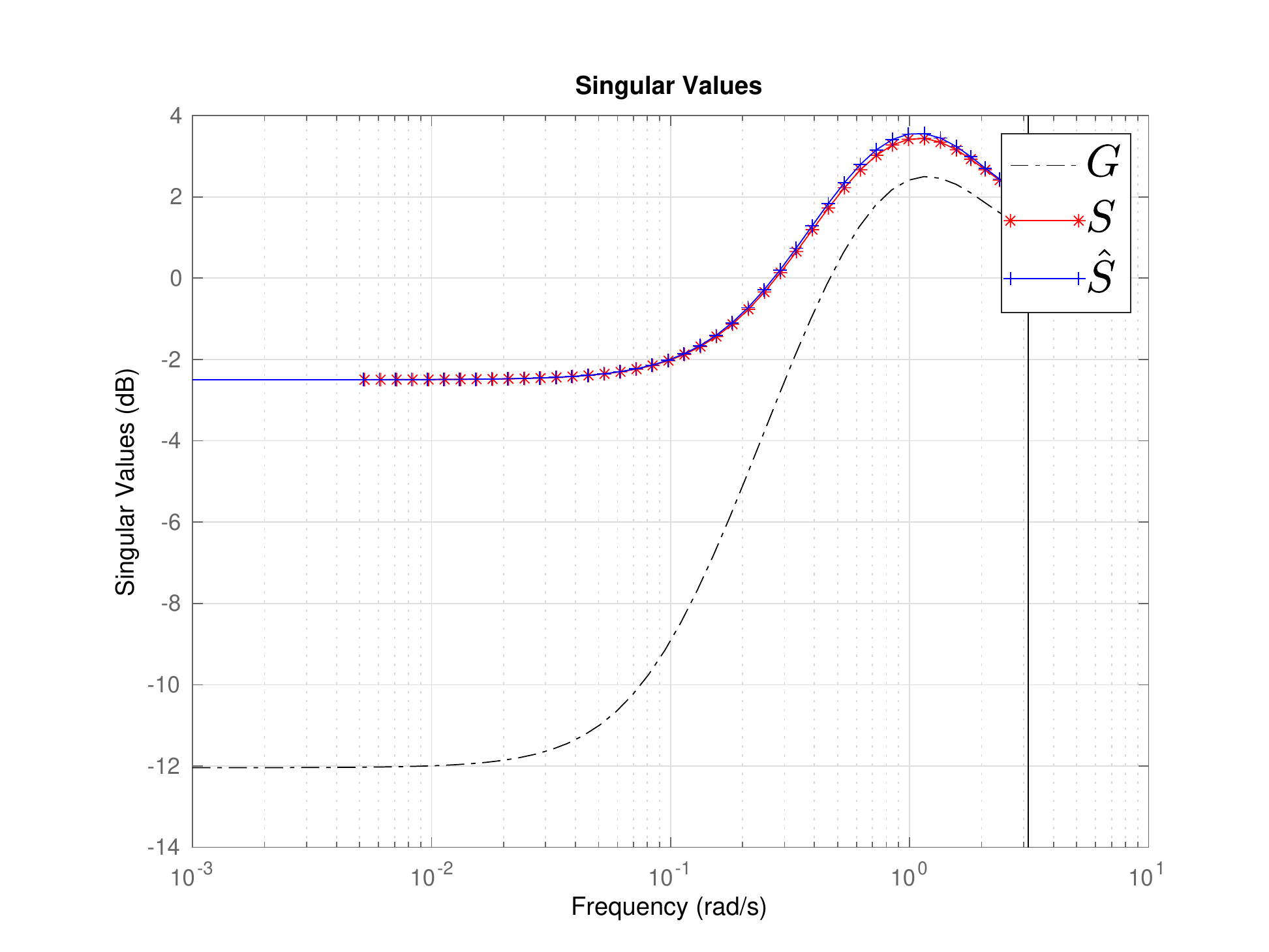}
    \caption{Singular values of $G,S$ and $\hat{S}$ along the unit circle.}
    \label{fig:bode}
\end{figure}

\begin{figure}
    \centering
    \includegraphics[width=8.4cm]{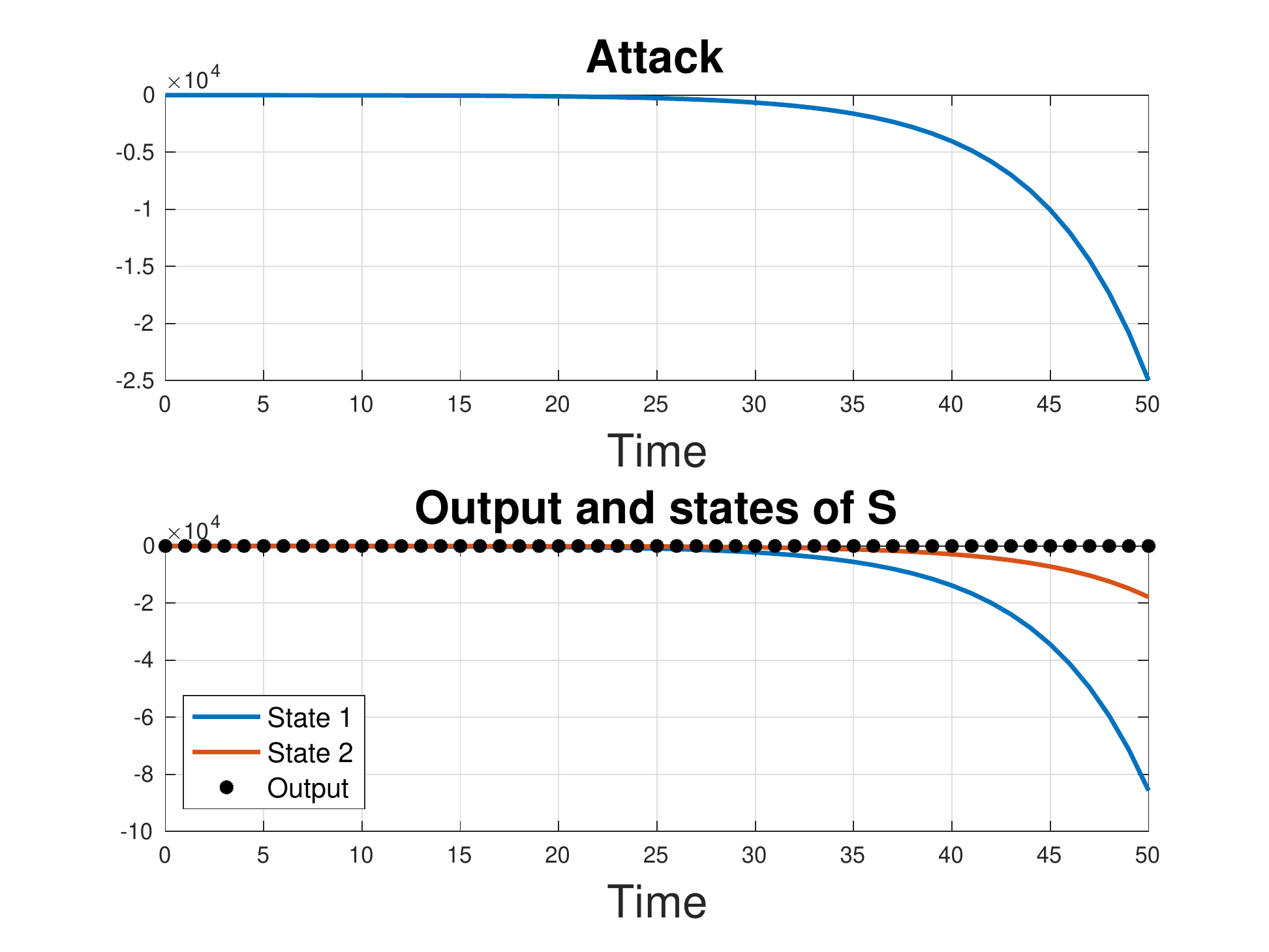}
    \caption{Performance of ZDA on $S(z)$}
    \label{fig:attack}
\end{figure}
\begin{figure}
    \centering
    \includegraphics[width=8.4cm]{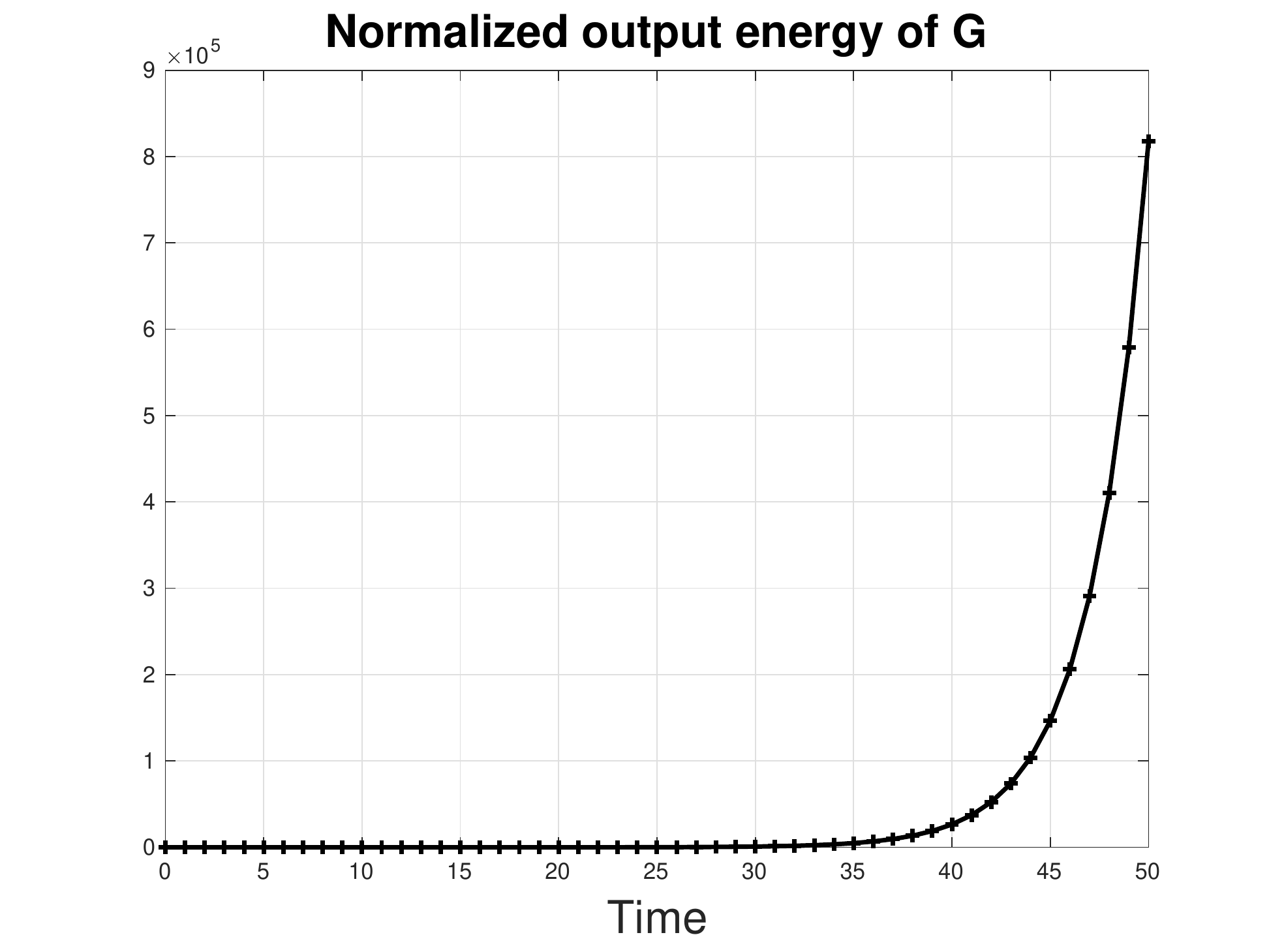}
    \caption{Output of $G(z)$ under ZDA}
    \label{fig:out_g}
\end{figure}

\end{exmp}

\section{Conclusion}\label{sec:Conclusion}
In this paper, we proposed a new architecture to enhance the privacy and security of NCS. We considered an adversary which first learns the plant dynamics, and then performs a ZDA. Under the proposed architecture, we show that it is possible to (i) introduce bias in the system knowledge of the adversary, and (ii) efficiently detect attacks. Through numerical simulations, we illustrate the efficacy of the proposed architecture. Future works include developing a systematic design procedure for $S$.

\bibliography{main}          

\end{document}